\documentclass[journal=jacsat,manuscript=article]{achemso}

\usepackage{chemformula} 
\usepackage[T1]{fontenc} 
\usepackage[version=3]{mhchem} 
\usepackage{natbib}
\usepackage{graphicx}
\usepackage{bm}
\usepackage{color,soul}
\usepackage{hyperref}
\usepackage{float}
\usepackage{color,soul}
\usepackage{algorithm2e}
\hypersetup{colorlinks=true, urlcolor=blue, citecolor=blue}
\usepackage{threeparttable}
\usepackage{mathtools}
\SectionNumbersOn
\setcitestyle{square}
\author{Shashi B. Mishra}
\author{B. R. K. Nanda}
\email{nandab@iitm.ac.in}
\affiliation{Condensed Matter Theory and Computational Lab, Department of Physics, \\ Indian Institute of Technology Madras, Chennai - 36, India}
\date{\today}

\title[An \textsf{achemso} demo]
  {Facet Dependent Catalytic Activities of Anatase TiO$_2$ for CO$_2$ Adsorption and Conversion }
  
\begin{document}


\begin{abstract}
 Understanding the atomic-scale interaction mechanism of CO$_2$ and H$_2$O on TiO$_2$ surface is crucial to establish a correlation between the catalytic efficiency with its exposed facet. Here, with the aid of a three-state model, nudged elastic band simulations, and DFT calculations, we examine the chemical restructuring of these molecules during the process of adsorption, coadsorption and conversion on (001) including (1$\times$4)-reconstructed, (010), and (101) facets of anatase TiO$_2$ and thereby, evaluate the step selective reactivity order. In addition, the results reveal the unexplored non-trivialities in the reaction mechanisms. For the most stable (101) facet, we show that the unfavorable carbonate complex formation becomes favorable by switching the reaction from endothermic to exothermic in the presence of water. Further, we find that the small binding energy does not necessarily imply physisorption. It can also give rise to chemisorption, where loss in energy due to repulsive Hartree and Madelung interactions is comparable to the energy gained through the chemical bonding. Such a scenario is demonstrated for the  CO$_2$ adsorption on (010) and (101) facets. Though (001) remains the most reactive surface, if it undergoes reconstruction, which happens at ultra high vacuum and high temperature, the number of active sites is reduced by three-fourth.
\end{abstract}
\section{\label{sec:level1}Introduction}
 The shape and size of semiconducting nanocrystals are the governing parameters towards enhancing their photocatalytic applications ~\citep{Barnard2005,Bai2017,Liu2011,Butburee2019,ARROUVEL2004,Tumuluri2017}. Owing to the greater stability, non-toxicity, natural earth abundance (low cost), and suitable band edge positions with respect to redox reactions, anatase TiO$_2$ (a-TiO$_2$) is established as the most preferred choice for photocatalytic and photovoltaic applications~\citep{Hadjiivanov1996,Diebold2003,Xu2011,Habi2013,Angelis2014}. However, the reactivity of a-TiO$_2$ is mostly dependent on the surface atomic distribution of the exposed crystal facet for the molecular adsorption processes ~\citep{Diebold2003,Hadjiivanov1996}. After the successful synthesis of single crystal anatase TiO$_2$ with a larger percentage of high surface energy (001) facet ~\citep{Yang2008,Liu2011,Han2009,Dai2009}, a wider attention has been paid towards the tailored synthesis of other high energy facets, such as (010), (111), (100), and (110) ~\citep{Pan2011,Xu2013,Ye2014,Xu2013a,Wen2011,Low2017,Zhou2011,Zhao2011,Pan2012,Chen2015,Gordan2012,Tumuluri2017}. Therefore, understanding the mechanism of adsorption and in turn the reactivity order of these low-index facets in comparison with the naturally stable low energy (101) facet is crucial to engineer high CO$_2$ adsorption and conversion efficiency on TiO$_2$ surfaces.

 The reactivity of different facets for a given adsorbate depends on the surface atomic distribution which determines the arrangement and coordination of active sites for the adsorption and conversion processes~\citep{Diebold2003,Hadjiivanov1996,Peng2017}. However, there are a divergence of views on the reactivity order for the low index facets of a-TiO$_2$. The theoretical studies based on adsorption of H$_{2}$O and other molecules such as methanol and formic acid on a-TiO$_2$ surface have reported that, (001) surface is more reactive than (101)~\citep{Vittadini1998,Gong2005}. Experimentally, it has been observed that, whenever synthesis leads to higher percentage of formation of (001) surface compared to (101), there is an enhancement in the photocatalytic activities such as reduction of CO$_2$ to methane, or degradation of organic contaminants ~\citep{Yang2008,Han2009,Liu2011,Dai2009,Yang2009,Yu2014,Wu2008}. But opposite reactivity order was proposed by few others. Pan \textit{et al.} observed that (010) shows highest photocatalytic reduction of CO$_2$ followed by (101) and (001) facets ~\citep{Pan2011}. This inference is seconded by other reports in the context of hydrogenation of CO$_2$ for H$_2$ generation ~\citep{Gordan2012,Xu2013,Xu2013a,Zhao2011}, and for the conversion of CO$_2$ to CH$_4$ in dry phase ~\citep{Ye2014}. In aqueous environment, while examining the degradation of dye molecule, they reported the reactivity order to be (001) $>$ (101) $>$ (010)~\citep{Ye2014}. It is suggested that in dry phase the catalytic activity is governed by the conduction band edge position, whereas in the aqueous solution the separation efficiency of the photo-generated charge carriers become the deterministic factor. Amidst these reports, a mechanism at the atomistic level correlating the reactivity with the band edge positions and the surface atomic distribution of the atoms is still absent.
 
 Although several theoretical reports have discussed the CO$_2$ adsorption and coadsorption with H$_2$O on (001) and (101) a-TiO$_2$ surface ~\citep{Indrakanti2008,He2010,Mino2014,Ma2016,Sorescu2011,Ji2016,Mishra2018}, a comparative study showing the reactivity order and its correlation with the surface electronic structures have not been presented. Also, the theoretical study of CO$_2$ adsorption on a-TiO$_2$ (010) surface remains least explored~\citep{Tumuluri2017}. Furthermore, due to higher surface energy of (001) facet, it undergoes 1$\times$4 reconstruction at ultrahigh vacuum and higher temperature ~\citep{Herman2000,Liang2001,DU2012,Sun2018,Silly2004}. The reactivities of the reconstructed (001)-(1$\times$4) surface needs to be understood further for its practical implications ~\citep{DeBenedetti2018,Wang2013,Shi2017,Shi2017a,Xu2017}.
 
So far the research have been focused towards CO$_2$ adsorption and conversion on a-TiO$_2$ (001) surface~\citep{Mino2014,Ma2016,Mishra2018}. The CO$_2$ adsorption on the reconstructed (001)-(1$\times$4) surface have not been explored so far. In particular, only limited reports are there for water, formic acid and methanol adsorption on the (001)-(1$\times$4) surface ~\citep{Gong2006,Selcuk2013,Xiong2016,Beinik2018,Liu2012}. It is reported that the water and methanol molecules dissociatively adsorb on the ridge positions, while they are weakly adsorbed at terrace sites. If this is true for any functional molecule adsorption, then the fraction of active sites are reduced ($\sim$ 0.25) as compared to the unreconstructed (001) surface~\citep{Gong2006,Beinik2018}. This implies that with (001)-(1$\times$4) reconstruction, the surface energy is reduced which stabilizes the surface and at the same time largely reduces its reactivity. Hence, it is suggested that with reconstruction, the reactivity order of (001) and (101) might reverse~\citep{Ye2014}. However, in a recent study, Feng \textit{et al.}, through a combined experimental and theoretical analysis of methanol adsorption and conversion to dimethyl ether(DME), have established that reconstructed (001)‐(1$\times$4) surface is more reactive than the (101) surface~\citep{Xiong2016}.
 
 Therefore, the intent of this work is to carry out a theoretical analysis on complete interaction mechanism starting from adsorption of CO$_2$ to conversion process for evaluation of the reactivity order among the (001), (010) and (101) surfaces and reconstructed (1$\times$4)-(001) surface. To achieve the objective, we have analyzed the surface electronic structure, L\"{o}wdin charges, and binding energies. Furthermore, hypothetical structures are investigated to construct a three state quantum model which is capable of demonstrating the adsorption mechanism and reactivity order.

 The comprehensive analysis presented in this work allows us to revisit the definition of physisorption and chemisorption through binding energy. The widely accepted assumption that weak binding implies physisorption is reexamined. We predict even weak binding can give rise to chemisorption in which, charge redistribution between the surface and adsorbate is such that the loss in the energy due to repulsive Hartree and Madelung interactions is comparable to the energy gained by the chemical bonding.
 
 The rest of the paper is organized as follows. The computational approaches and structural designs are presented in section 2 and the results are analyzed in section 3. Here, in subsection 3.1, we have discussed about the surface atomic distribution and electronic structure of a-TiO$_2$(001)-(1$\times$1), reconstructed (001)-(1$\times$4), (010) and (101) surfaces. In 3.2 and 3.3, we  have presented the CO$_2$ adsorption on TiO$_2$ facets and explained the interaction mechanism through a three-state model and charge density analysis. In 3.4, we explore the coadsorption configurations of CO$_2 - $H$_2$O complexes. In 3.5, we have examined the formation of bicarbonate complex. The summary and conclusions are provided in section 4.

\section{Computational Methods}
 The density functional calculations are performed using plane-wave basis sets as implemented in Quantum ESPRESSO~\citep{Giannozzi_2017}. The exchange-correlation potential is approximated through the Perdew-Burke-Ernzerhof general gradient approximation (PBE-GGA) functional~\citep{Perdew1996}. The ultrasoft pseudopotentials are used, in which the valence states of Ti include 12 electrons in 3s, 3p, 3d and 4s shells, while O include 6 electrons and C include 4 electrons in 2s, and 2p shells. The kinetic energy cutoff to fix the number of plane waves is taken as 30 Ry, while the cutoff for the augmented electron density is set to be 300 Ry. The dispersion corrections have been included through the semi-empirical Grimme-D2 van der Waals correction~\citep{Grimme2006}. A $4\times4\times1$ k-mesh is considered for the structural relaxation, while for electronic structure calculations, a denser k-mesh of $8\times8\times1$ is used. The convergence criterion for self-consistent energy is taken to be 10$^{-6}$ Ry. The structural relaxations are performed until the force on each atom is lower than 0.025 eV/\AA. The L\"{o}wdin charges on individual atoms are calculated by subtracting the electronic charge (by summing the contribution of valence orbitals of the corresponding atom on each occupied) from the total number of valence electrons of the corresponding atom. The charge density difference plots for CO$_2$ adsorbed TiO$_2$ surfaces are calculated by subtracting the charge density of the TiO$_2$-CO$_2$ system from that of pristine TiO$_2$ and CO$_2$ in the same coordinate space as that of the interacted one. The structural and charge-density plots are generated using the visualization tool VESTA~\citep{Momma2008}. The minimum energy path for migration of hydrogen atom during the formation of surface bicarbonate (HCO$_3$) is calculated using the transition state theory based climbing image nudged elastic band (CI-NEB) approximation~\citep{Henkelman2000}.
 
 For anatase, the calculated bulk lattice parameters are a = 3.794 \AA, c = 9.754 \AA{} which agree well with the previously reported experimental and theoretical values ~\citep{Burdett1987,Mo1995,Lazzeri2001,Lopez2016,Liu2012,Zhao2010}. The surfaces are constructed using a bi-dimensional slab model with a vacuum of 15 \AA{}. Except the (1$\times$4) reconstructed (001) surface, all other surfaces, i.e., (001), (010) and (101) have been constructed by considering equal number of TiO$_2$ units (i.e. 24 units). The anatase (001) surface was modelled using a p(3$\times$2) supercell (11.38$\times$7.59\AA$^2$) with a four TiO$_2$ layers ($\sim$ 8.1\AA{} thick). The (010) surface was modelled with a p(1$\times$2) supercell (9.65$\times$7.59\AA$^2$) with a six layered TiO$_2$ ($\sim$ 9.5\AA{} thick), while for (101) a p(1$\times$2) surface slab (10.37$\times$7.59\AA$^2$) with a three layered thick TiO$_2$ unit ($\sim$ 9.4\AA) is used. For (001)-(1$\times$4) reconstructed surface, we have considered the widely adopted and experimentally verified ADM model~\citep{Lazzeri2001a} in which the bridging O$_{2f}$ atoms along [100] are replaced with rows of TiO$_3$ units in a 1$\times$4 periodicity, with a four layered thick TiO$_2$ slab in a p(2$\times$4) supercell (15.18$\times$7.59\AA$^2$) having 34 TiO$_2$ units. A slab of the aforementioned thickness is reported to be sufficient for molecular calculations~\citep{Gong2005,Shi2017,Gong2006,Vitale2018,Lopez2016}. For clean anatase (001), (010) and (101) surfaces, all the atoms are fully relaxed, whereas in the case of (001)-(1$\times$4), the atoms in the bottom layer are fixed to their bulk positions while all other atoms are allowed to move freely.
 
The surface energies of (001), (010) and (101) facets are calculated by considering a 1$\times$1 unit cell with a vacuum of 15 \AA. The surface energy ($E_s$) is estimated by using the following expression ~\citep{Gong2005}.
     \begin{equation}\label{eq1}
         E_S = \frac{1}{A}(E_{slab} - N\times E_{bulk})
     \end{equation}
where, $E_{slab}$ and $E_{bulk}$ represent the total energy of the slab and the total energy per formula unit of bulk TiO$_2$ respectively, while N stands for the number of TiO$_2$ formula units present in the slab, and A is the total exposed area including top and bottom surfaces of the slab. In the case of clean (1$\times$1) TiO$_2$ (001), (010) and (101) surfaces, the top and bottom layers are equivalent having 4, 9 and 9 TiO$_2$ formula units. While for the case of (1×4)-(001), identical terminations are made on the top and bottom surfaces with 18 TiO$_2$ formula units as has been carried out by Gong et al.~\citep{Gong2005,Gong2006}.  A 4$\times$4$\times$1 k-mesh is used for the (001), (010) and (101) facets, while for the (1×4)-(001) a k-mesh of 4$\times$1$\times$1 is used. During relaxation, all the atoms are allowed to move freely until a force convergence of $\sim$ 0.025 eV/\AA{} is achieved. We estimated the surface energy of (001) to be 0.98 J/m$^2$ with an error of $\pm$0.03 J/m$^2$. For the (101) surface, $E_S$ is estimated to be 0.40$\pm$0.05 J/m$^2$ and for the (010) the value is 0.50$\pm$0.02 J/m$^2$. For reconstructed (001)-(1×4), $E_S$ is calculated to be 0.50$\pm$0.01 J/m$^2$. The surface energy calculations are in good agreement with the previous reports~\citep{Gong2005,Gong2006,Lazzeri2001,Vitale2018,Angelis2014,Zhao2010,Yang2008}.

 The binding energy of functional molecules are calculated using the following equation
 \begin{equation}\label{eq2}
    BE = E_{{TiO_2}/molecule} - E_{TiO_2} - E_{molecule}
 \end{equation}
where, $E_{{TiO_2}/molecule}$ and $E_{TiO_2}$ represent the total energies of optimized adsorbed and pristine surfaces. The $E_{molecule}$ represents the total energy of isolated molecule (CO$_2$, H$_2$O) which is calculated by keeping the molecule in a cubic box of size 20 \AA.

To calculate the vacuum reference in each of surface, we have calculated the planar average of the electrostatic potential ($V^{PA}$) and the macroscopic average ($V^{MA}$) of each of these surfaces. The $V^{PA}$ is obtained by averaging the raw three dimensional potential ($V^{raw}$) using the following expression~\citep{Samanta2018}
 \begin{align} \label{eq3}
 V^{PA}(z)=\frac{1}{S}\int_sV^{raw}(x,y,z)dxdy,\nonumber\\
 V^{MA}(z)=\frac{1}{c}\int^{z+c/2}_{z-c/2}V^{PA}(z^\prime)dz^\prime.
 \end{align}
Where, $S$ is the area of each slab and $c$ is the length of one period. For (001), (001)-(1$\times$4), (010) and (101) slabs, the  $c$ values are 4.28, 4.28, 3.82 and 3.55 \AA{}, respectively. 

\section{Results and Discussion}
\subsection{Electronic structure of bare surfaces}
\begin{figure}[hbt!]
    \centering
    \includegraphics[width=16.5cm, height=11.2cm]{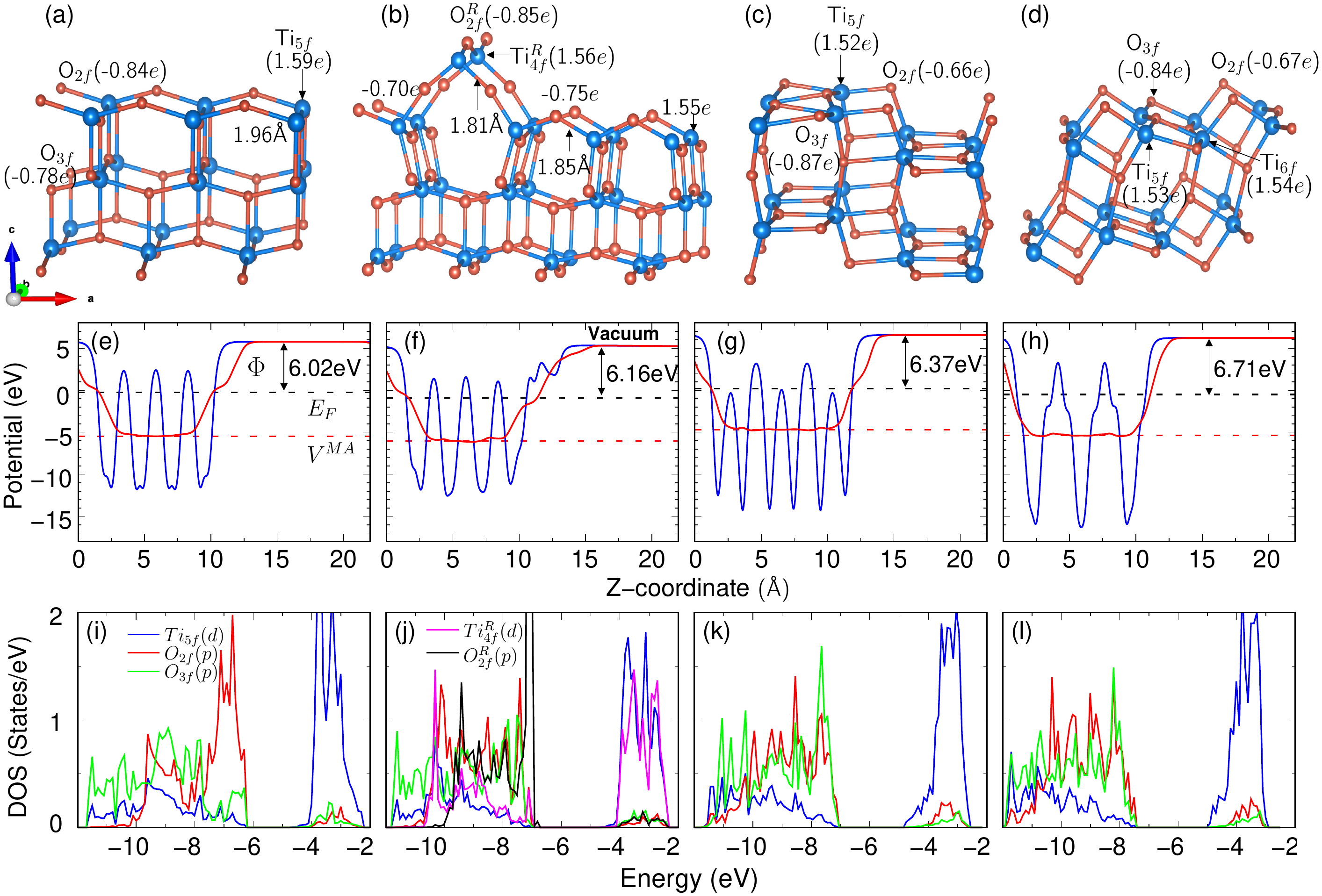}
    \caption{(a-d) The optimized structure of a-TiO$_2$ (001)-(1$\times$1), (1$\times$4) reconstructed (001), (010) and (101) surfaces, respectively. The L\"{o}wdin charges on specific surface atoms Ti$_{4f}^R$, Ti$_{5f}$, O$_{2f}$, O$_{2f}^R$ and O$_{3f}$ are indicated. (e-h) Their corresponding electrostatic potential profiles depicting the vacuum level ($V^{vac}$) and planar average ($V^{PA}$) potential. The macroscopic average ($V^{MA}$) potential, for each of the surfaces, calculated using Eq. \ref{eq3} are shown in red lines. The Fermi level ($E_F$) is shown as black horizontal dashed line. The work function ($\Phi$) is estimated by subtracting $E_F$ from the vacuum level ($V^{vac}$). (i-l) The partial densities of states (DOS) of Ti$_{5f}$, O$_{2f}$, and O$_{3f}$ atoms for these surfaces along with the ridge atoms Ti$_{4f}^R$ and O$_{2f}^R$ in reconstructed (001)-(1$\times$4) surface. The partial density of states are plotted relative to vacuum level.}
    \label{fig1}
\end{figure}

 To have an understanding of the reactivity order among the clean a-TiO$_2$ (001) and reconstructed (001)-(1$\times$4), (010), and (101) facets, the first step is to analyze the atomic arrangements of the relaxed pristine surfaces (see Figure \ref{fig1} (a-d)), and thereafter their corresponding electrostatic potential profiles and electronic structures. As shown in Figure \ref{fig1} (a), the pristine (001)-(1$\times$1) surface creates a zigzag chain of Ti$_{5f}-$O$_{2f}-$Ti$_{5f}$ along [100] with the oxygen terminated surface, while along [010], the chain follows the pattern of Ti$_{5f}-$O$_{3f}-$Ti$_{5f}$ ~\citep{Lazzeri2001,Angelis2014,Lopez2016}. After relaxation, the Ti$_{5f}$ atoms are slightly pushed inwards, whereas O$_{2f}$ atoms move upwards. The Ti$-$O$_{2f}$ bond length increased from the bulk value of 1.94 to 1.96 \AA{}, and the Ti$_{5f}-$O$_{2f}-$Ti$_{5f}$ angle reduced from 155.34\textdegree{} to 151.04\textdegree, whereas the O$_{3f}-$Ti$_{5f}-$O$_{3f}$ angle increased by 2.15\textdegree. The axial Ti$_{5f}-$O$_{3f}$ bond length is reduced by 0.06 \AA{} and the equatorial Ti$_{5f}$ -- $O_{3f}$ distance changes negligibly ($\sim$ 0.007\AA). In the case of (001)-1$\times$4 reconstructed surface~\citep{Lazzeri2001a}, the ridges appear above the regular terrace sites which contains two-fold coordinated O$_{2f}^R$ and four-fold coordinated Ti$_{4f}^R$ atoms (see Figure \ref{fig1} (b)). After geometry optimization, Ti$_{5f}-$O$_{2f}$ bond length is reduced to around $\sim$ 1.81$-$1.85 \AA{} from that of relaxed (1$\times$1) surface which is in good agreement with the earlier report~\citep{Lazzeri2001a}. This results in change in charge values on Ti$_{5f}$ and O$_{2f}$ atoms on the terrace atoms as compared to that of (001)-(1$\times$1) surface. The charge on terrace Ti$_{5f}$ is reduced from +1.59 to +1.55$e$, while on O$_{2f}$ it is reduced from $\sim$ -0.84 to -0.72$e$. On the ridge site, the charges on Ti$_{4f}^R$ and O$_{2f}^R$ are +1.56$e$ and -0.85$e$, respectively.
 
 The atomic arrangement for the (010) surface~\citep{Angelis2014} is shown in Figure \ref{fig1} (c). After relaxation, the Ti$_{5f}$ atoms are slightly pushed inwards, as a result, the planar angle of O$_{2f}-$Ti$_{5f}-$O$_{2f}$ is reduced to 164.98\textdegree. Also, the bond distance between Ti$_{5f}-$Ti$_{6f}$ is decreased from 3.07 to 2.86 \AA. Similarly, the Ti$_{5f}-$O$_{2f}$ and Ti$_{5f}-$O$_{3f}$ bond lengths are reduced to 1.83 and 1.79 \AA, respectively. Similarly, the (101) surface  exposes bulk Ti$_{6f}$ and O$_{3f}$ atoms along with the Ti$_{5f}$ and O$_{2f}$ atoms ~\citep{Vittadini1998,Liu2019,Sorescu2011} (Figure \ref{fig1}(d)). With the relaxation, the surface Ti$_{5f}$ and Ti$_{6f}$ atoms are pushed inwards, and the separation between Ti$_{5f}$ and Ti$_{6f}$ is reduced to 2.84 \AA. The Ti$_{5f}-$O$_{2f}$ and Ti$_{5f}-$O$_{3f}$ bond lengths decreased to 1.78 and 1.83 \AA{} respectively from ideal value of 1.94 \AA. Also, the Ti$_{6f}-$O$_{2f}$ decreases to 1.85 \AA, whereas Ti$_{6f}-$O$_{3f}$ increases to 2.0 \AA{} from 1.94 \AA.
 
 To estimate the surface potential, we have plotted the planar ($V^{PA}$) and macroscopic average ($V^{MA}$) potential for each facets (see computational details) in Figure \ref{fig1} (e-h). The vacuum level is estimated to be 5.76, 5.23, 6.55 and 6.21 eV for (001)-(1$\times$1), (001)-(1$\times$4), (010) and (101) facets, respectively. The work function, $\phi$ (=vacuum - $E_F$) for these facets are calculated to be 6.02, 6.16, 6.37 and 6.71 eV, respectively (see Figure \ref{fig1} (e-h)). On the basis of surface atomic arrangements, (001) surface is found to be more reactive for the adsorbates as compared to (101) and (010) facets ~\citep{Gong2005,Chen2015a,Vittadini1998,Beinik2018,Vitale2018}. 
 
 The intricate details of the reactivity of the facets towards the adsorbates can be explained by examining the surface electronic structure. In Figure \ref{fig1} (i-l), we have plotted the partial density of states (DOS) of Ti$_{5f}$-$d$, O$_{3f}$-$p$ and O$_{2f}$-$p$ atoms for all the considered facets. In addition, for (001)-(1$\times$4) surface, the ridge Ti$_{4f}^R$-$d$ and O$_{2f}^R$-$p$ DOS are also shown. While the relative band edge positions qualitatively agree with the experimental reports obtained using XPS and optical absorption spectra ~\citep{Pan2011, Ye2014,Xu2013,Xu2011}, they quantitatively differ by as much as $\sim$0.8 eV for (001), (101) and (010) surfaces. From the DOS, we observed that the valence band maximum (VBM) corresponding to the O-$p$ states of (001) surface lies nearly 0.9 and 1.4 eV higher in energy as compared to the (010) and (101) surfaces, respectively. Our prediction of VBM and CBM for (001) and (101) surfaces agree very well with the earlier theoretical reports ~\citep{Chen2015a,Chamtouri2017}. With reconstruction, the O$_{2f}^R$-$p$ and Ti$_{4f}^R$-$d$ populates the VBM and CBM of the (001) surface which are also now shifted to resemble to that of the (101) and (010) surfaces (see Figure \ref{fig1}(j)).

 The electronic structure of (001) surface is distinct from the (010) and (101). For the former, the O$_{2f}$-$p$ DOS are separated from the O$_{3f}$-$p$ DOS and lie higher in energy, and have a narrow bandwidth. Whereas for the (010) and (101) facets, there is no noticeable difference among the O$_{3f}$ and O$_{2f}$ states. Since, the less diffused surface states forming the VBM can easily form chemical bonding with the molecules during adsorption as compared to that of diffuse surface states, the (001) surface is more reactive towards the functional molecules~\citep{Chen2015a}. This is in agreement with the earlier theoretical report that for water and formic acid, (001) surface is observed to be more reactive as compared to that of (101) surface ~\citep{Vittadini1998,Gong2005,Gong2006}. The bandwidth of O$_{2f}$/O$_{3f}$-$p$ states in (010) and (101) surfaces are similar and hence, their reactivity towards functional molecules will be closely related. For (001)-(1$\times$4) reconstructed facet, the ridge O$_{2f}^R$-$p$ states differ from that of terrace O$_{2f}$-$p$ states implying that the O$_{2f}^R$ is active towards adsorbates whereas terrace O$_{2f}$ becomes passive after reconstruction~\citep{Gong2006,Beinik2018}.
 
\subsection{CO$_2$ adsorption on TiO$_2$ surfaces}
\begin{figure}
    \centering
    \includegraphics[width=14.5cm, height=8cm]{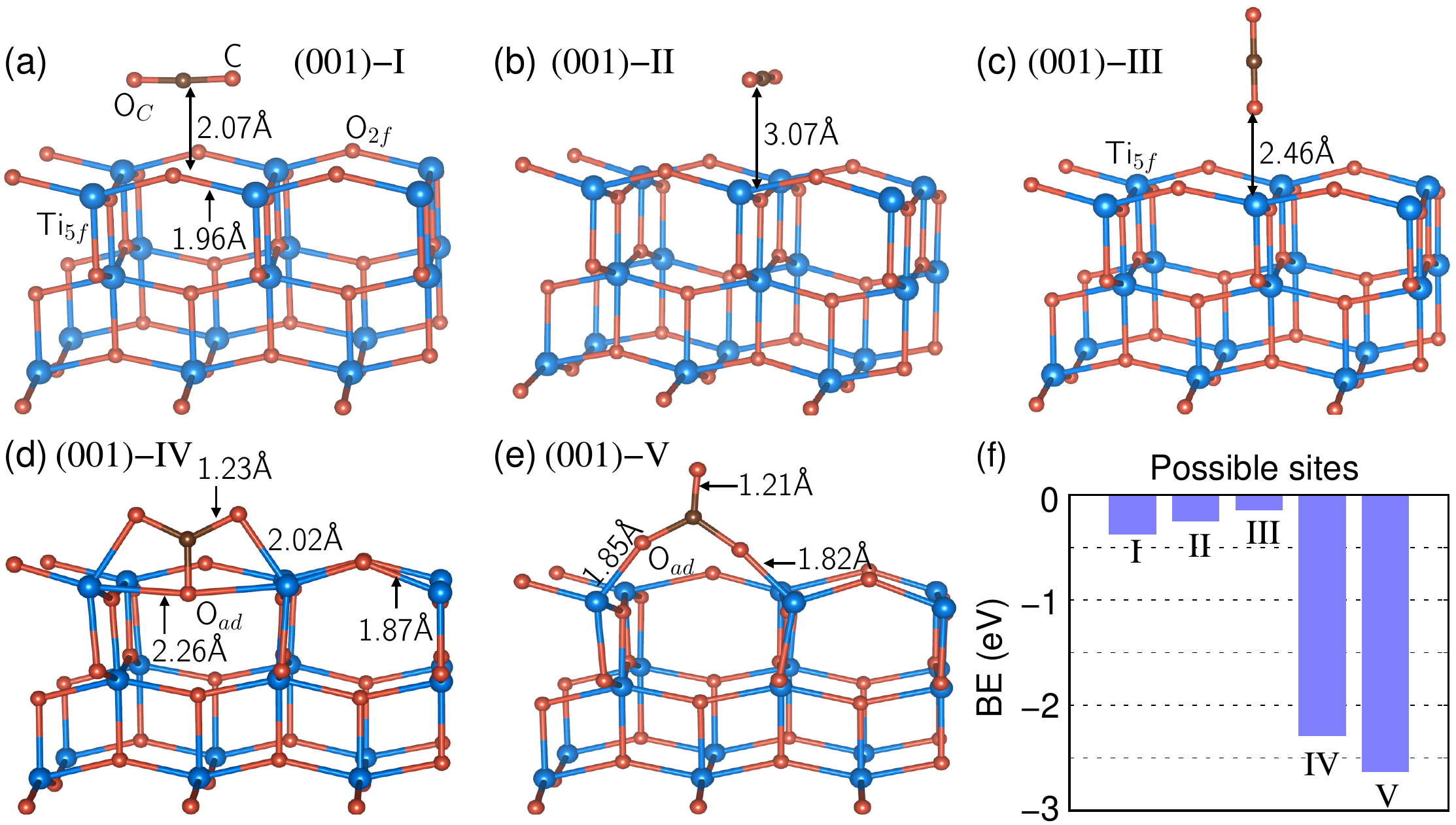}
    \caption{(a-e) The optimized structure of CO$_2$ adsorbed on TiO$_2$ (001) surface with adsorption occurring at various sites. Here, we have considered one CO$_2$ molecule on $3\times2$ (001) slab of four-layered thick with a coverage of $\sim$ 16.6 \%. (f) The binding energy (BE) corresponding to these configurations.}
    \label{fig2}
\end{figure}

\subsubsection{(001) surface}
 The adsorption of CO$_2$ on a-TiO$_2$ (001)-(1$\times$1) surface was discussed in detail in our previous studies~\citep{Mishra2018}, but for comparison purposes to other surfaces, we have summarized these calculations for a $3\times2$ slab model as shown in Figure \ref{fig2}. As discussed in our earlier work, the CO$_2$ molecule is weakly adsorbed over ontop O$_{3f}$, Ti$_{5f}$ and hollow positions (Figure \ref{fig2}(a-c)), and  retains its linear configuration. When CO$_2$ is adsorbed on top of O$_{2f}$ site, the structure undergoes deformation, and a carbonate complex is formed (Figures \ref{fig2}(d-e)) which is in agreement with the previous reports~\citep{Indrakanti2008,Mino2014}. This is consistent with the anticipation made for the (001)-(1$\times$1) bare surface electronic structure, where the VB predominates with the O$_{2f}$-$p$ states and hence a chemical bonding between ontop O$_{2f}$ and the adsorbate is expected. The binding energies shown in Figure \ref{fig2}(f) indeed shows that the carbonate complex formation, configurations (001)-IV and (001)-V, lead to stronger binding between the adsorbate and adsorbent. In this study, we will explore further the adsorption process in these two configurations. 
\begin{figure}
    \centering
    \includegraphics[width=16cm, height=8.0cm]{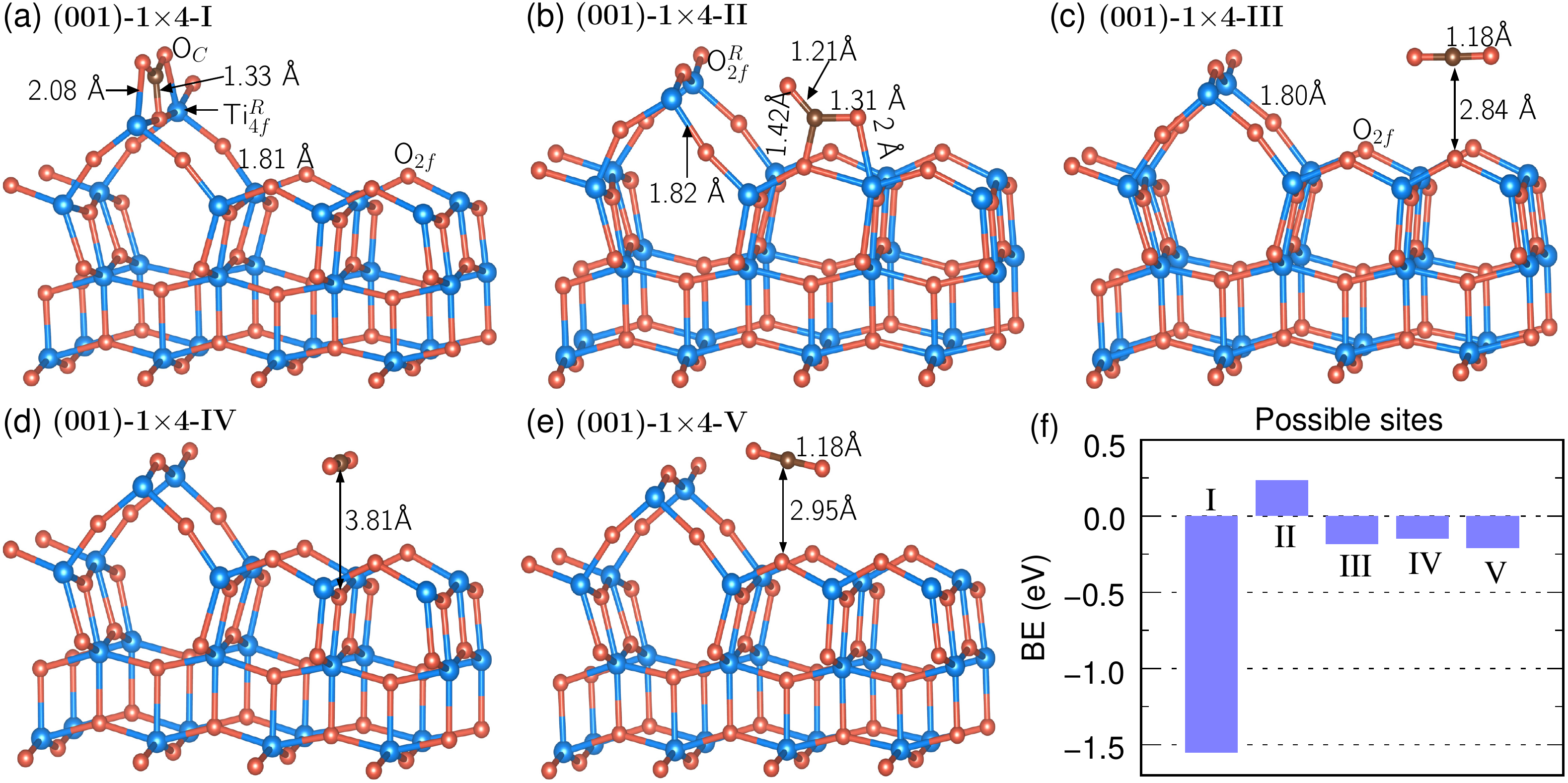}
    \caption {The optimized structure of CO$_2$ adsorbed on a-TiO$_2$ (001)-(1$\times$4) reconstructed  surface with adsorption occurring at various sites. (f) The binding energy (BE) corresponding to these configurations.}
    \label{001-1x4}
\end{figure}

\subsubsection{(001)-(1$\times$4) reconstructed surface}
 We have examined the CO$_2$ adsorption at five possible sites of (001)-(1$\times$4) surface as shown in Figure \ref{001-1x4}. When CO$_2$ is adsorbed at the $O_{2f}^R$ site, a carbonate complex (CO$_3^{\delta-}$) with a O$_C-$C$-$O$_C$ bond angle of 130\textdegree{} is formed (see Figure \ref{001-1x4}(a)). The complex is formed with an increase in the C$-$O$_C$ bond length to 1.32 \AA{} and by a new chemical bond between C and $O_{2f}^R$ with a bond length of 1.33 \AA{}. The final outcome is a chemisorption of CO$_2$ which is also reflected from its binding energy plot (Figure \ref{001-1x4}(f)).

 As Figure \ref{001-1x4}(b) shows when CO$_2$ is adsorbed in the bent configuration at the terrace O$_{2f}$ site, the C$-$O$_{2f}$ bond length becomes 1.42 \AA{} while Ti$_{5f}-$O$_{C}$ separation remains at 2 \AA. As a result of this interaction, the O$_C-$C$-$O$_C$ angle becomes 131.8\textdegree{} and an asymmetric carbonate complex is formed. Although chemical restructuring take place, the positive binding energy implies this process to be endothermic which is discussed in detail in section 3.3. For adsorption at other terrace sites, the CO$_2$ retains its linear geometry (see Figure \ref{001-1x4}(c-e)) with weak binding energies implying physisorption. Hence, due to (1$\times$4) reconstruction, the terrace O$_{2f}$ sites become passive towards the CO$_2$ molecule. This resonates with the earlier report on adsorption of water and formic acid  on (001)-(1$\times$4) surface ~\citep{Beinik2018,Gong2006}. The passivity of O$_{2f}$ atoms are most likely due to shortening of terrace Ti$_{5f}-$O$_{2f}$ bonds ($\sim$ 1.81$-$1.85 \AA)~\citep{Lazzeri2001a,Gong2006}.

\subsubsection{(010) surface}
\begin{figure}
    \centering
    \includegraphics[width=13cm, height=7.5cm]{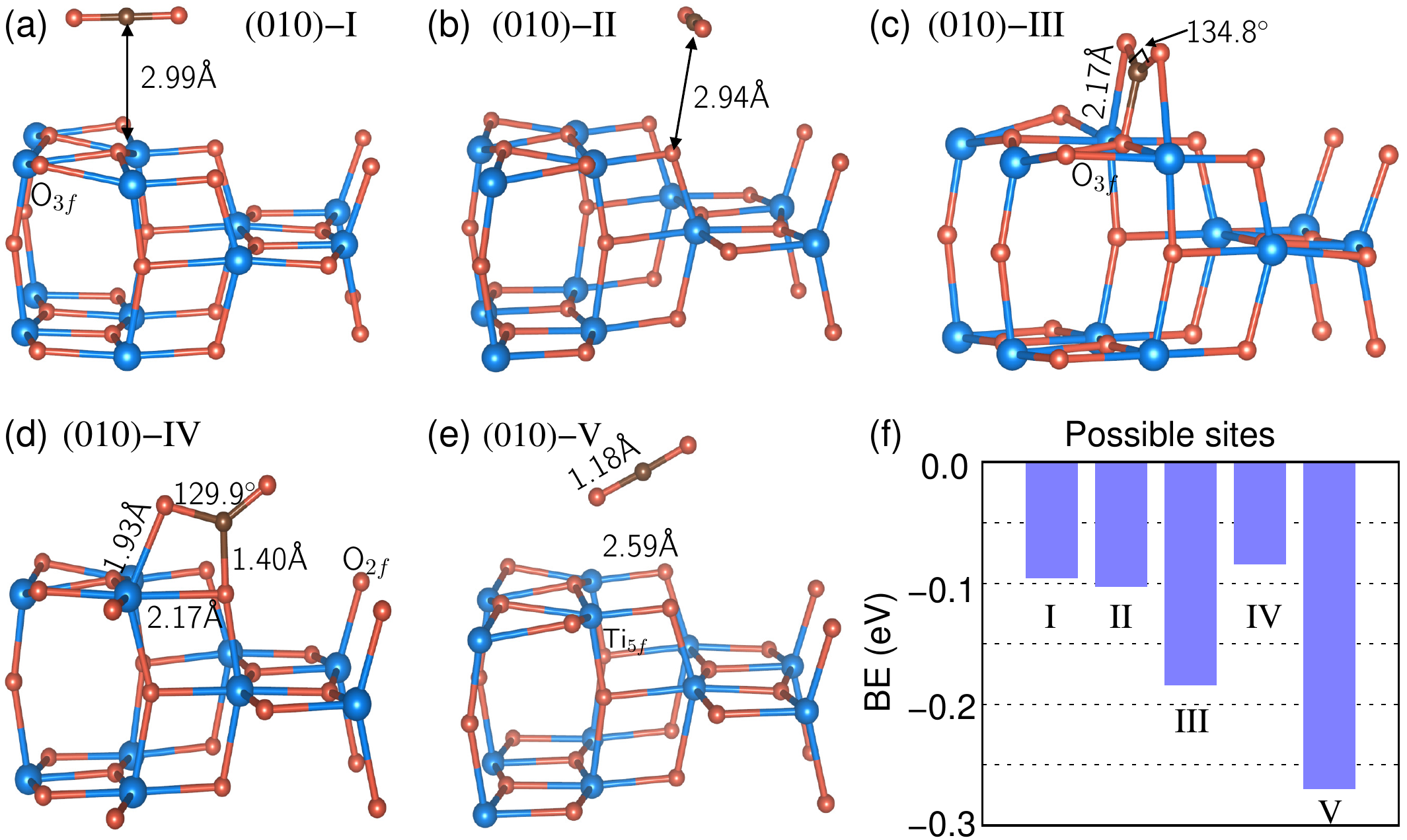}
    \caption{(a-e) The optimized structure for CO$_2$ adsorbed configurations on (010) surface. (f) The binding energy (BE) corresponding to these configurations.}
    \label{fig3}
\end{figure}

 The optimized structures of various possible adsorption motifs of CO$_2$ on the (010) surface are shown in Figure \ref{fig3} (a-e) and their corresponding binding energies in Figure \ref{fig3}(f). For configurations I and II, the CO$_2$ retains its linear geometry and lies at a distance $\sim$ 2.95 \AA{} from the top of surface layer. When CO$_2$ is placed on top of O$_{3f}$ site, the C atom is pushed downward, and it forms a bond with the surface O$_{3f}$ atom at a bond length of 1.41 \AA{} and the each Ti$_{5f}-$O$_{C}$ separation at 2.17 \AA. The C$-$O$_{C}$ bond length is increased to 1.26 \AA{} and the O$_{C}-$C$-$O$_{C}$ bond angle reduced to 134.78\textdegree. This resembles the formation of a tridentate carbonate complex (CO$_3^{\delta-}$) as shown in Figure \ref{fig3}(c).
 
 A similar carbonate complex is formed when the adsorption takes place at the surface O$_{2f}$ site as shown in Figure \ref{fig3}(d). In this configuration, the O$_{2f}-$C bond length is 1.40 \AA{} and one of the O$_C$ atoms forms a bond with surface Ti$_{5f}$ at a separation of 1.93 \AA. Also, one of the C$-$O$_C$ bonds (O$_C$ bonded with Ti$_{5f}$) is elongated to 1.33 \AA, while the other C$-$O$_C$ bond remains at 1.20 \AA. The O$_C-$C$-$O$_C$ bond angle is 129.95\textdegree{} and O$_{2f}-$C$-$O$_C$ bond angle becomes 105.8\textdegree. However, the binding strength is enhanced when the CO$_2$ molecule is adsorbed above the Ti$_{5f}$ site in a tilted configuration. The CO$_2$ retains linear geometry and lies at a bond separation of 2.59 \AA{} from the Ti$_{5f}$ site. Compared to the (001) surface, the BE is very weak ($<$ 0.3 eV) for all the possible configurations in this facet (see Figure \ref{fig3}(f)). However, among these configurations, although BE is lower for the (010)-III and (010)-IV configurations as compared to (010)-V, stronger chemical interactions between surface and adsorbate takes place in the former cases leading to chemisorption process which is discussed in more detail in the subsection 3.3.
 
\subsubsection{(101) surface}
\begin{figure}
    \centering
    \includegraphics[width=14cm, height=8cm]{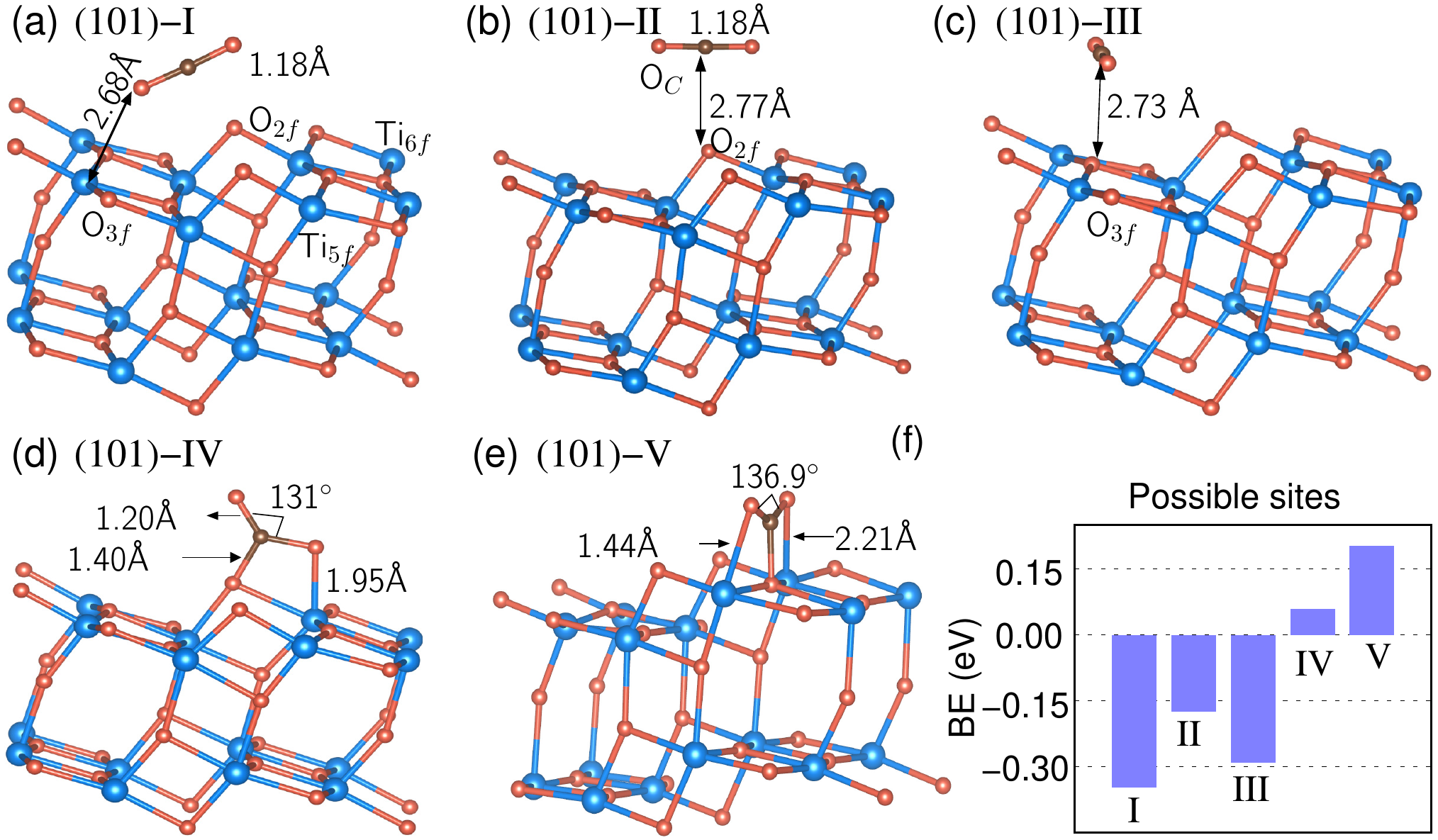}
    \caption{The optimized structure for CO$_2$ adsorption on (101) surface at various configurations. (f) The binding energy (BE) corresponding to these configurations.}
    \label{fig4}
\end{figure}
 The optimized structure of CO$_2$ interaction at five possible adsorption motifs on (101) surface are displayed in Figure \ref{fig4}(a-e). Akin to the case of (010) surface, the binding energies are weak (see Figure \ref{fig4}(f)) which was anticipated from their electronic structure (Figure \ref{fig1} (k) and (l)). The BE plot shows that the strongest adsorption on the (101) surface happens when CO$_2$ is in tilted position above the Ti$_{5f}$ site at a vertical separation of 2.68 \AA. This observation is in agreement with the earlier reports, where they mentioned the tilted configuration of CO$_2$ to be energetically most preferred on TiO$_2$ (101) surface ~\citep{Sorescu2011,Mino2014,Ji2016,Ma2016}. The CO$_2$ retains its linear geometry in most of the cases except configurations IV and V (see Figure \ref{fig4}(d) and (e)), where, it undergoes structural deformation forming bidentate and tridentate carbonate like complexes respectively. Interestingly, the BE for formation of these complexes is calculated to be endothermic even though they involve stronger chemical bonding between the surface and adsorbate. In subsection 3.4 we will see that these complex formations become exothermic in the presence of H$_2$O.

\subsection{Mechanism of CO$_2$ adsorption}
\subsubsection{Three-state model}
 \begin{figure}
    \centering
    \includegraphics[width=17cm, height=16cm]{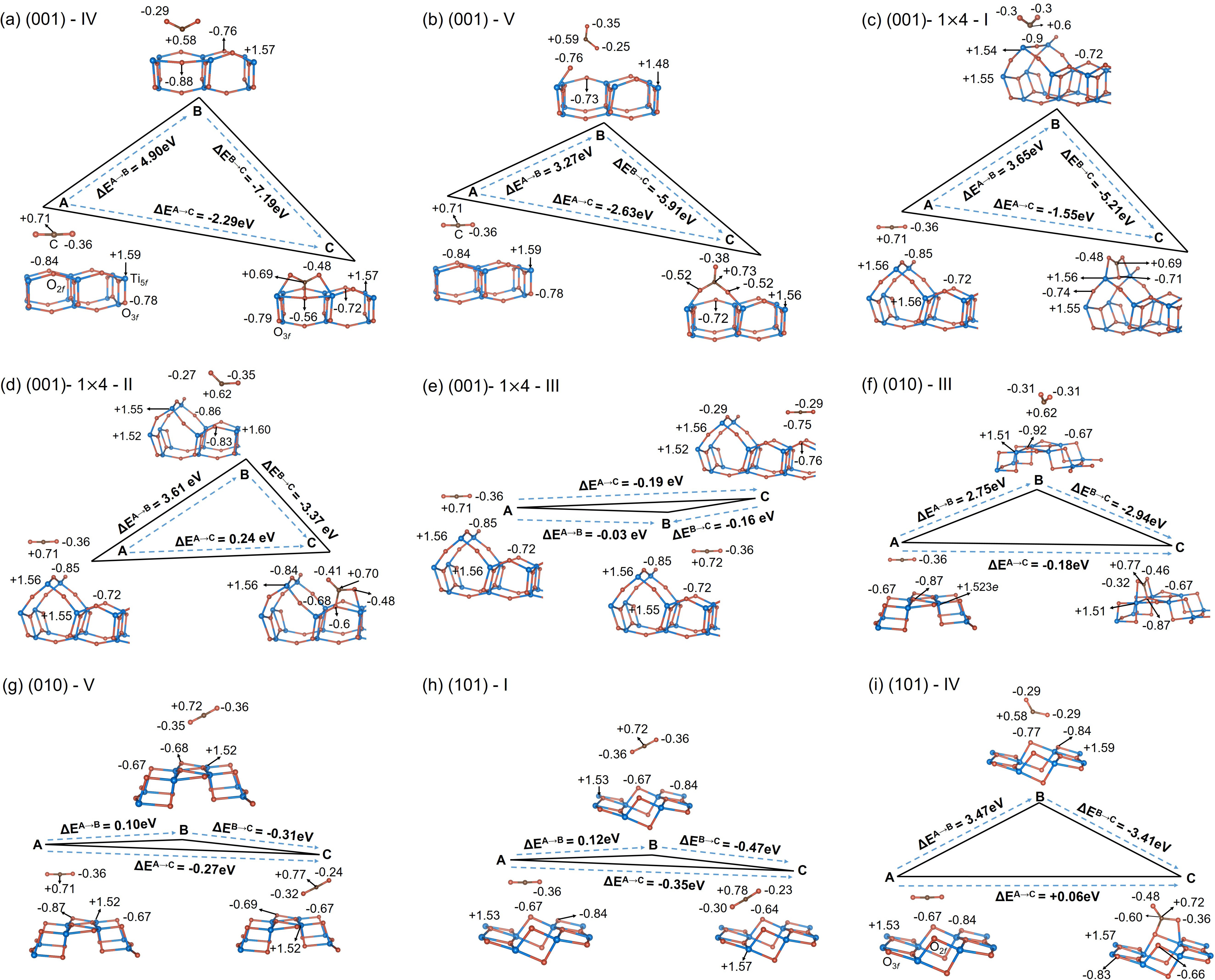}
    \caption{The three-state model explaining the CO$_2$ adsorption mechanism on relevant configurations of a-TiO$_2$ (001) and (1$\times$4) reconstructed (001), (010), and (101) surfaces. States-A and C represent the situation before and after adsorption respectively. The relative energies as well as site-specific L\"{o}wdin charges for each state are mentioned. State-B is a geometrical replica of state-C, but without any interaction between the molecule and surface. The charges are in electron ($e$) units.}
    \label{three-state}
\end{figure}
 To further analyze the mechanism of CO$_2$ adsorption, we have now carried out L\"{o}wdin charge analysis for relevant adsorption configurations of each facets. These are IV and V for unreconstructed(001); I, II and III of (1$\times$4) reconstructed (001); III and V of (010); and I and IV of (101) surface. The strength of adsorptions for each of these cases are presented through a three state model ~\citep{Mishra2018} which are schematically illustrated in Figure \ref{three-state}. In this model, states-A and C represent the situation before and after adsorption, respectively. However, state-B which is hypothetical, is a geometrical replica of state-C, but without any interaction between the adsorbate and adsorbent. The charges at each site and the relative energies for all three states corresponding to the aforementioned adsorption configurations are listed in the Figure \ref{three-state}. This model suggests that, the strength of adsorption is directly proportional to the instability occurred to the adsorbate and adsorbent for the hypothetical state of B and the site charge differences between state-A and state-C. For the case of (001) surface, both the instability and the site charge differences are large, and therefore, the adsorption is stronger leading to a strong binding energy. 

 In the case of (001)-(1$\times$4)-III, (010)-V and (101)-I configurations, neither the instability is large, nor there is a significant charge transfer. Therefore, the adsorbate and adsorbent are weakly interacting and hence, these are the cases of physisorption. For the configurations (001)-(1$\times$4)-II, (010)-III and (101)-IV, the site charge differences, i.e charge on each of the mentioned specific sites such as O$_{2f}$, Ti$_{5f}$, and O$_{C}$, etc. before and after adsorption, and the energy difference between state-A and C are small suggesting that these states are nearly degenerate. However, the instability of state-B is large and since the site charge differences between state-B and C in these configurations are significant, a chemisorption cannot be ruled out which will be clear from the following subsection.
 
\subsubsection{Analysis of Charge Density}
 \begin{figure}[hbt!]
    \centering
    \includegraphics[width=12cm, height=9.5cm]{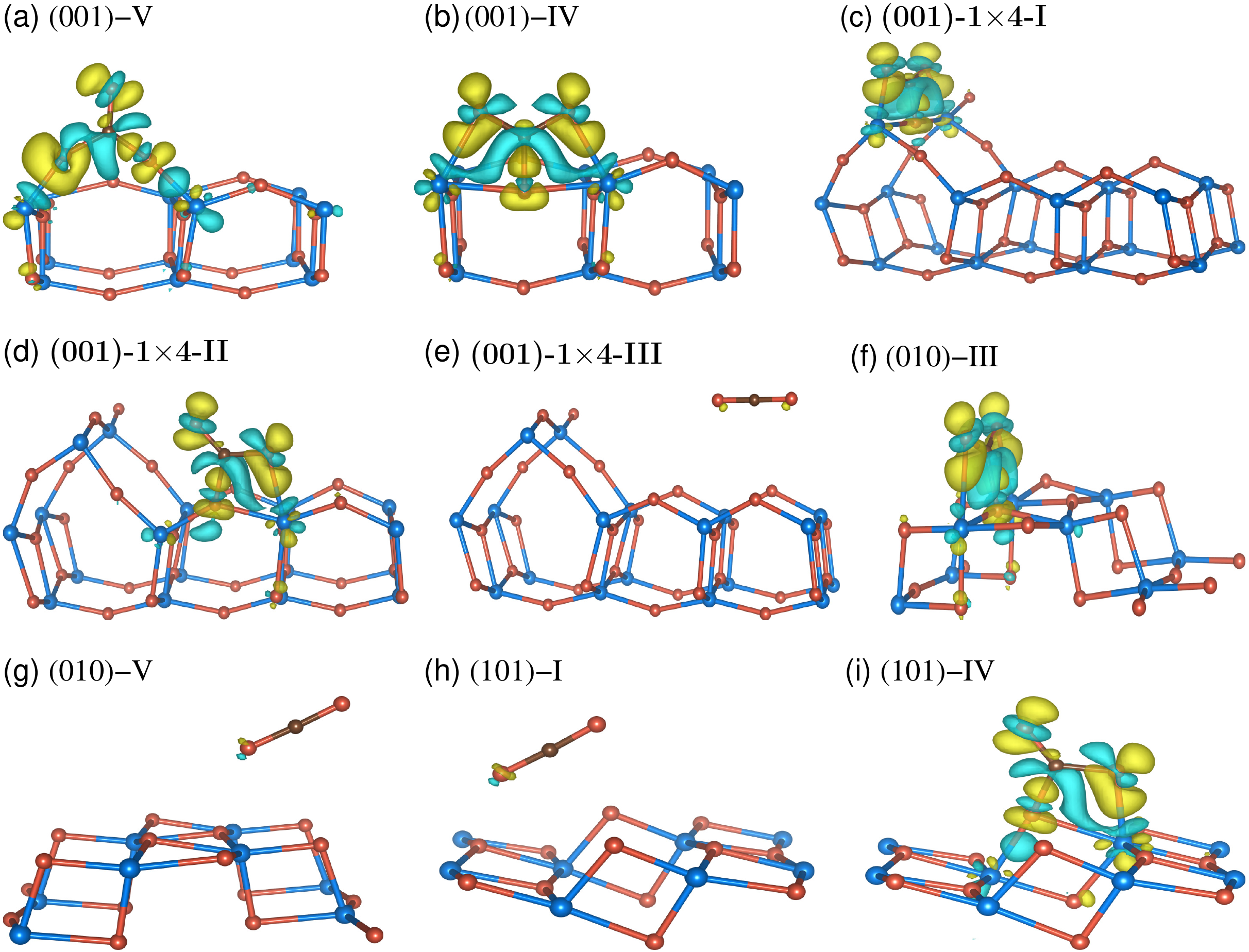}
    \caption{The chemical restructuring of CO$_2$ adsorbed on TiO$_2$ (001) and (1$\times$4) reconstructed (001), (010) and (101) surfaces. (a-i) The charge density difference ($Q^C -$ $Q^B$ of the three-state model, see Figure \ref{three-state}). The yellow and cyan charge contours represent charge accumulation and depletion regions, respectively with an iso-value of 0.009 e/\AA$^3$. The absence of charge density difference and retaining of molecular eigenstates of CO$_2$ for (001)-1$\times$4-III, (010)-V and (101)-I configurations imply physisorption. For the remaining, the CO$_2$ states overlap with surface states and the charge contours indicate chemisorption process.}
    \label{fig6}
\end{figure}
 The chemical interactions are further examined with the aid of charge densities for each of the configurations of Figure \ref{three-state} and the results are shown in Figure \ref{fig6}. We have plotted the difference in the charge densities of state-B and state-C of the aforementioned three state model. We gather that for the cases of (001)-(1$\times$4)-III, (010)-V and (101)-I, the electron sharing between the molecule and the surface is negligible suggesting no chemical bonding and therefore, it reconfirms the physisorption as predicted from the three state model. For the cases of (001)-(1$\times$4)-I, (001)-IV and V, the electron sharing is significant resulting into a strong chemical bonding between the surface and adsorbate. This chemical bonding is responsible for a large and negative binding energy which is typical of a chemisorption process. 
 
 The cases of (001)-(1$\times$4)-II, (010)-III and (101)-IV are non-trivial. The binding energy is weak and can be endothermic (e.g. (001)-(1$\times$4)-II, and (101)-IV) and at the same time, the electron sharing is as significant as in the case of (001)-(1$\times$4)-I, 001-IV and V, which suggests that the adsorbate and adsorbent favours chemical bonding.
 
\begin{figure}[hbt!]
    \centering
    \includegraphics[width=8.5cm, height=4.5cm]{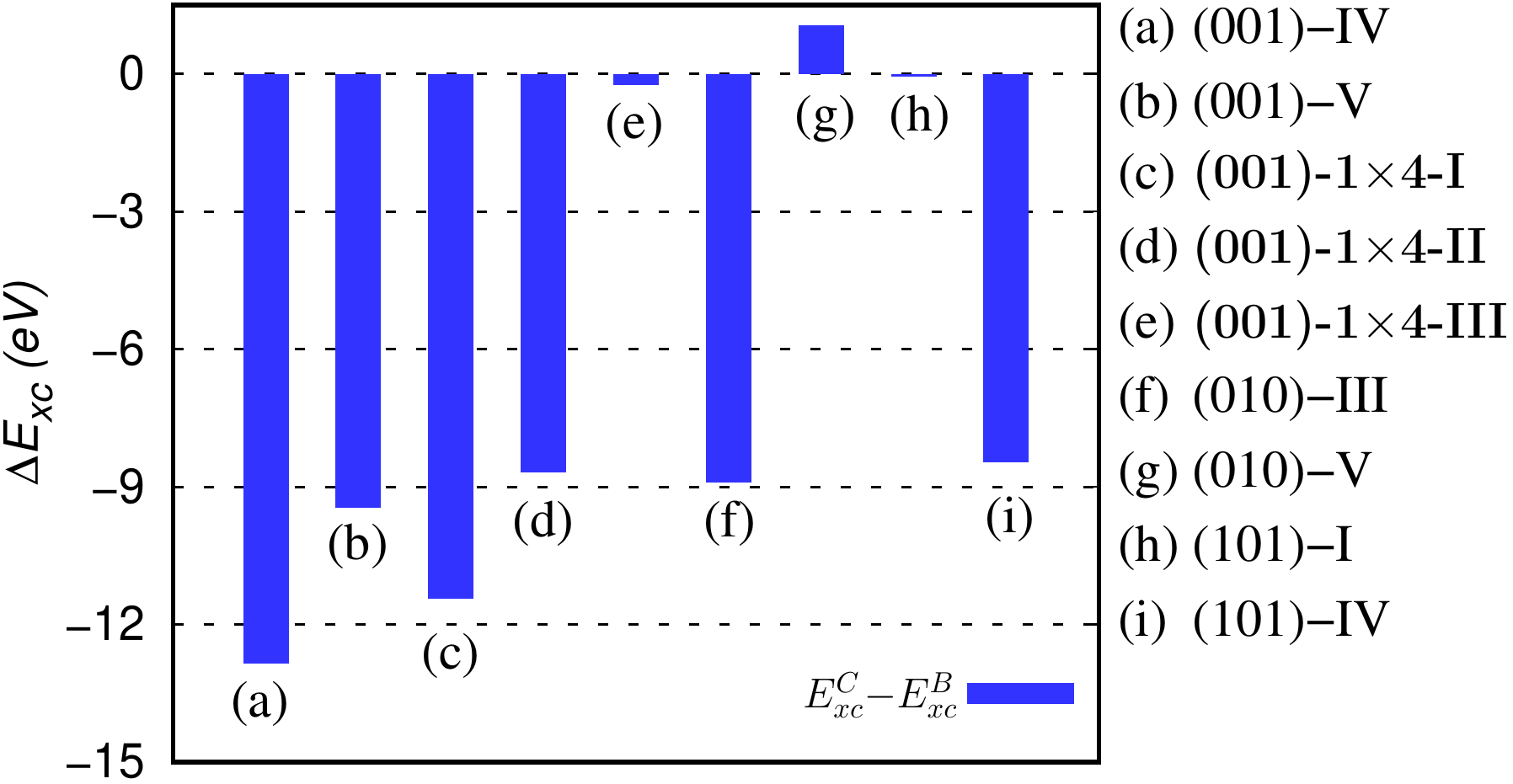}
    \caption{The difference in exchange-correlation (XC) energy between state B and state C for the three-state model configurations of (001)-IV, V; (001)-(1$\times$4)-I, II, III; 010-III, V; and 101-I, IV.}
    \label{xc}
\end{figure}
 
 A quantitative measure of the chemical interaction between the adsorbate and substrate can be obtained by examining the relative change in the net exchange-correlation (XC) energy as one moves from state-B to state-C of the three state model. With the charge cloud deformations, the contributions from all types of interactions add to the total energy change. However, the change in the exchange-correlation energy, which has a non-linear dependence on the charge density, is a better indicator of the chemical interaction between two entities. In fact in a recent study, it is shown that the change in the exchange energy can be directly mapped to a chemical bonding initiated between two isolated species~\citep{Naik2017}.
 
 From the results shown in Figure \ref{xc}, we gather that change is significant for (001)-(1$\times$4)-I, (001)-IV and (001)-V suggesting the larger electron participation in the chemical bonding. For the cases of (001)-(1$\times$4)-II, (010)-III and (101)-IV, though the binding energy is weak (0.23 eV, -0.18 eV and 0.06 eV), the change in XC is significant indicating stronger chemical interactions between the molecule and the surface. Same cannot be presumed for the cases of (001)-(1$\times$4)-III, (010)-V and (101)-I, where the change in the XC is negligible as well as the binding energy (-0.18 eV, -0.27 eV and -0.35 eV) is weak and therefore these three configurations offer physisorption. 

 This paradox of having weak binding and at the same time the molecule is chemisorbed can be understood by proposing the following hypothesis. For these non-trivial cases, the chemical bonding redistributes the charge in such a way that the loss in the repulsive interaction arising from Hartree and Madelung (ion-ion) is comparable to the  gain the energy via this chemical bonding.

\subsection{Coadsorption of H$_2$O and CO$_2$ on TiO$_2$ surfaces}
\begin{figure}
    \centering
    \includegraphics[width=17cm, height=3.5cm]{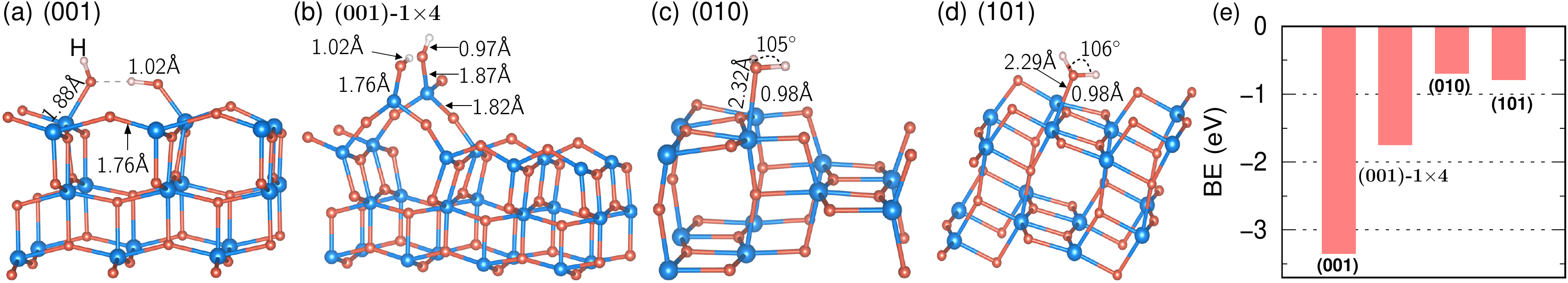}
    \caption{(a-d) The optimized structure of H$_2$O adsorption on (001) and 1$\times$4 reconstructed (001), (010) and (101) facets and (e) their corresponding binding energies. On (001) surface, the H$_2$O is dissociated to hydrodxyl (-OH) groups, whereas on (001)-(1$\times$4), it dissociates to -OH groups only at ridge positions. While H$_2$O remains undissociated on (010) and (101) surfaces.}
    \label{h20}
\end{figure}
 In all practical conditions, the CO$_2$ adsorption occurs under aqueous environment. It is reported that the adsorption energies of CO$_2$ on TiO$_2$ surfaces are greatly affected by the presence of H$_2$O. Most importantly, the binding energy of CO$_2$ gets enhanced by $\sim$ 0.1$-$0.2 eV when coadsorbed with single H$_2$O~\citep{Sorescu2012,Yin2015,Klyukin2017}. Also, with coadsorption, CO$_2$ tends to form bicarbonate complexes which can further be processed for complete reduction. To understand the mechanism of coadsorption, it is prudent to first have a look at the H$_2$O interaction on TiO$_2$ (001) and (1$\times$4) reconstructed (001), (010) and (101) facets. In Figure \ref{h20} (a-d), we have shown the optimized ground state structure for H$_2$O adsorption on these facets. As computationally the adsorption is examined on a surface formed by a 3$\times$2 supercell, this represents a water coverage of $\sim$ 16.6\% over (001) surface. From the Figure \ref{h20} (a), we observe that, the H$_2$O molecule dissociates to two hydroxyl (-OH) groups on (001) surface. Similarly, the H$_2$O molecule is dissociated to hydroxyl ions when adsorbed at the ridge sites of (001)-(1$\times$4) surface (see Figure \ref{h20} (b)), whereas on terrace positions it remains in the molecular form (see Supplementary Information). As a result, the number of active sites after reconstruction is reduced by one-fourth. This observation is in agreement with the earlier reports~\citep{Selcuk2013,Beinik2018}. The H$_2$O is weakly adsorbed and remains in the molecular form on (010) and (101) surface (Figure \ref{h20} (c)and (d)). The BE plot shows that the molecular adsorption comes with weaker binding energy (Figure \ref{h20}(e)). Though the case of H$_2$O adsorption on (001) and (101) surfaces had been reported earlier~\citep{Vittadini1998,Agosta2017,Sorescu2012}, the case of adsorption on the (010) surface had not been examined before.
 
\begin{figure}
    \centering
    \includegraphics[width=16.cm, height=5.6cm]{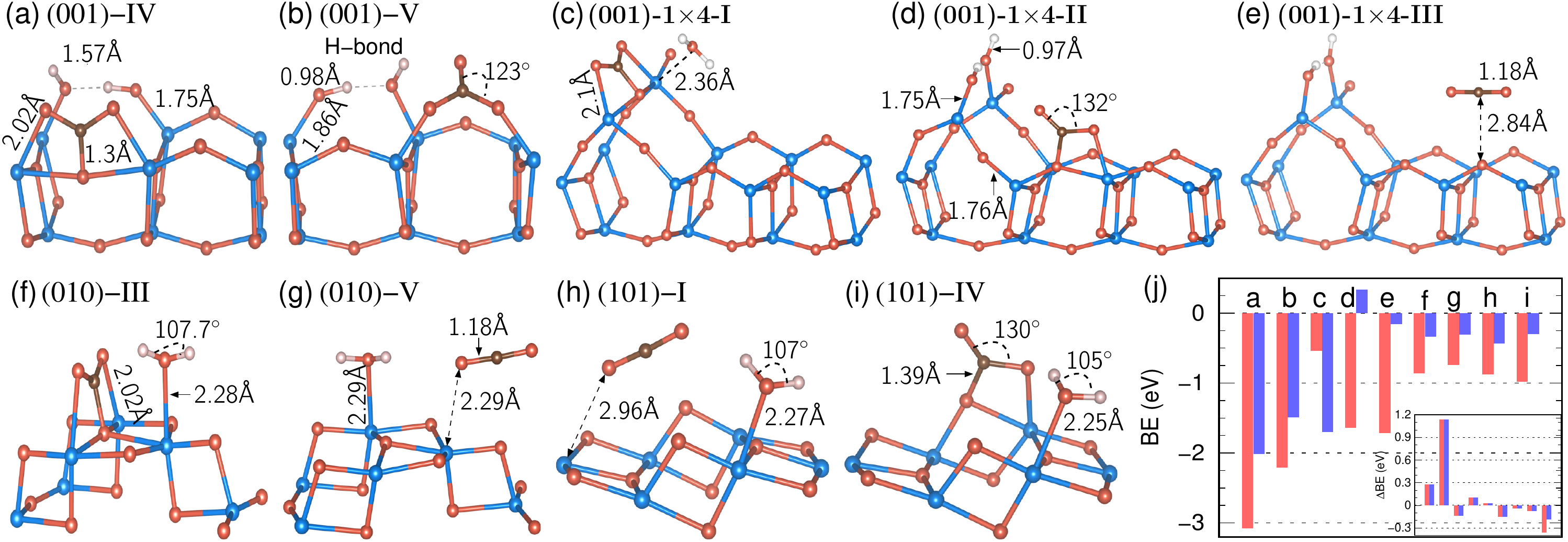}
    \caption{(a-i) The optimized CO$_2-$H$_2$O coadsorbed configurations on TiO$_2$ (001), (001)-(1$\times$4), (010) and (101) surfaces. (j) The BE corresponding to these configurations are calculated in two ways. The BE of CO$_2$ on H$_2$O adsorbed TiO$_2$ surface, and BE of H$_2$O on CO$_2$ adsorbed TiO$_2$ surface are shown in blue and red bars, respectively. The change in BE ($\Delta$BE) with respect to separate adsorptions of CO$_2$ and H$_2$O are shown in the inset.}
    \label{coads}
\end{figure}

 Now, we shall examine the case of coadsorption. Earlier, it has  been shown that the final configuration remains the same irrespective of the order of adsorption of CO$_2$ and H$_2$O, namely, CO$_2$ adsorption followed by H$_2$O, H$_2$O adsorption followed by CO$_2$ or simultaneous adsorption CO$_2$ and H$_2$O ~\citep{Mishra2018}. Here, we have adopted the case of H$_2$O adsorption on the CO$_2$ adsorbed surface. The final coadsorbed  configurations are shown in Figure \ref{coads}(a-i). By comparing these configurations with the isolated adsorptions (see Figures \ref{fig2}-\ref{fig4}, \ref{h20}), we infer that the coadsorption process is seeming to be simply addition of two isolated CO$_2$ and H$_2$O adsorptions. To bring an quantitative analysis of the coadsorption, in Figure \ref{coads}(j), for each of the configurations, we have shown two different binding energies, namely, the BE of CO$_2$ adsorption on the H$_2$O adsorbed TiO$_2$ surface (blue), and the BE of H$_2$O adsorption on CO$_2$ adsorbed TiO$_2$ surface (red). The change in the BE with respect to the respective adsorptions are shown in the inset of the figure. Here, we draw two interesting conclusions. Firstly, in the process of coadsorption, the binding strength of the H$_2$O and CO$_2$ decreases for the most reactive (001) surface and now the configuration (001)-IV becomes more preferable than (001)-V.  Secondly, for the cases of (010)-III and (101)-IV, where the CO$_2$ was chemisorbed on the bare TiO$_2$ surface despite very weak BE, the binding strength has increased in the presence of H$_2$O due to the formation of hydrogen bond. Most significantly, the reaction has turned exothermic from endothermic for the configuration (101)-IV which is also now the most preferred configuration for the (101) facet. On the basis of binding energies corresponding to coadsorption of CO$_2$ and H$_2$O molecules, we conclude that both (001)-1$\times$1 as well as (001)-1$\times$4 reconstructed surfaces are more reactive as compared to the (010) and (101) surfaces. The coadsorbed configurations can be classified into four categories. (a) CO$_2$ chemisorption and H$_2$O dissociation (001)-IV, (001)-V and (001)-(1$\times$4)-II. (b) CO$_2$ chemisorption and H$_2$O molecular adsorption ((001)-(1$\times$4)-I, (010)-III and (101)-IV). (c) CO$_2$ physisorption and H$_2$O molecular adsorption ((001)-(1$\times$4)-III, (010)-V and (101)-I). The last one is believed to be not conducive for further reduction of CO$_2$ into a bicarbonate complex ~\citep{Klyukin2017,Mishra2018}. 

\subsection{Formation of bicarbonate complex}
  \begin{figure}[hbt!]
    \centering
    \includegraphics[width=16.0cm, height=10cm]{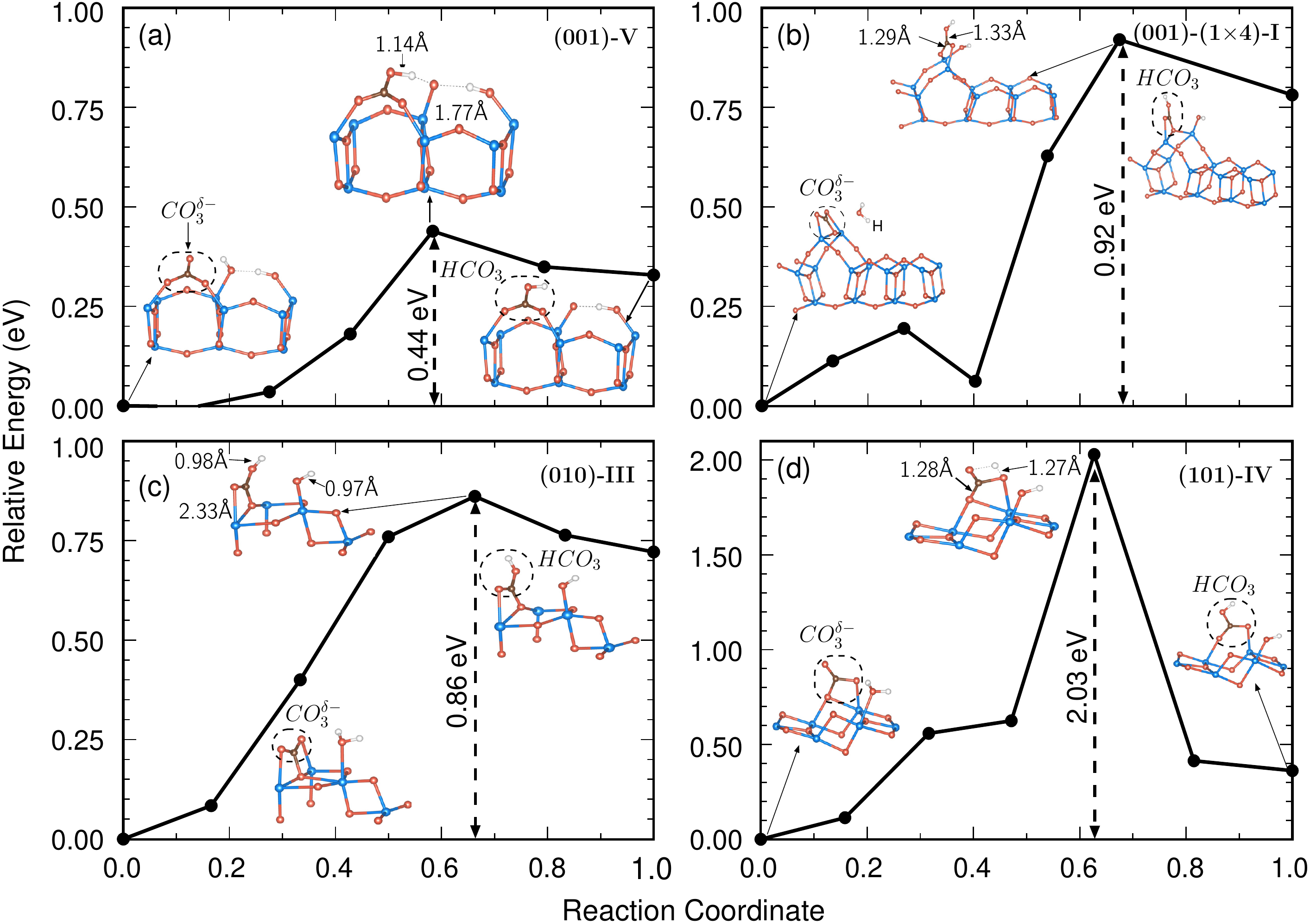}
    \caption{Minimum energy pathways during the process of conversion of surface carbonate to bicarbonate complex (see Eq. 5) for configurations: (a) (001)-V, (b) (001)-(1$\times$4)-I (reconstructed), (c) (010)-III and (d) (101)-IV. The structure corresponding to initial and final state along with the transition states (TS) are indicated. The TS structural coordinates are given in the Supplementary Information.}
    \label{fig10}
\end{figure}
The presence of water transforms the carbonate complex into a  bicarbonate complex formulated through the following chemical steps ~\citep{Ye2012,Mino2014,Sorescu2012}. 
  \begin{eqnarray} 
     TiO_2 + CO_2 &\longrightarrow& TiO-CO_3\\
     TiO_2-CO_3 + H_2O &\longrightarrow& TiO-(OH)-HCO_3
 \end{eqnarray}
 The transition pathway for the formation of bicarbonate complex on TiO$_2$ (001) and (1$\times$4) reconstructed (001), (010) and (101) surfaces are displayed in Figure \ref{fig10}(a-d). In the case of (001)-(1$\times$1) surface, the hydrogen ion detaches from one of the two hydroxyl (-OH) groups and forms a bond at a length of $\sim$ 1.14 \AA{} with the surface CO$_3^{\delta-}$ complex by overcoming an activation energy barrier of $\sim$ 0.44 eV. On (001)-(1$\times$4) (reconstructed), (010) and (101) surfaces, the hydrogen ion separates from the molecularly adsorbed H$_2$O, and moves towards the surface carbonate complex to form HCO$_3$ by overcoming energy barriers of 0.92, 0.86 and 2.03 eV respectively. We observed that the dissociative adsorption of H$_2$O lowers the activation energy barrier significantly. Based on the activation energy barriers for CO$_2$ reduction to HCO$_3$, it can be inferred (001) is most reactive followed by (010) and (101) surface. 

\section{Summary and Conclusion}
 In summary, in the quest of establishing the reactivity order, we have systematically examined the adsorption and coadsorption of CO$_2$ and H$_2$O, and the formation of HCO$_3$ on unreconstructed as well as (1$\times$4)-reconstructed (001), (010), and (101) facets of anatase TiO$_2$ using DFT calculations and NEB simulations. An insight to the CO$_2$ adsorption on these facets are provided through a three-state model.  The CO$_2$ adsorption on both (1$\times$1) and (1$\times$4) reconstructed clean anatase TiO$_2$ (001) surfaces leads to formation of tridentate carbonate complex with stronger binding energies. On the (001)-(1$\times$4) surface, the complex formation takes place at ridge position, whereas on the terrace position a bidentate carbonate complex is formed through an endothermic process. Similarly, the carbonate complex formation on (010) and (101) surface is energetically less favored. We predict that weak binding energy not necessarily implies physisorption as concurrently there may be a strong chemical bonding between the adsorbate and substrate through electron sharing, and a significant reduction in the binding energy through repulsive Hartree and Madelung interactions which was also observed from the charge density difference. Specific to the most stable (101) surface, we find that the presence of H$_2$O makes the CO$_2$ favour the carbonate complex formation. The present study reports that the change in the exchange correlation energy after adsorption can be treated as an indicator of the strength of chemisorption.
 
 From quantitative analysis of H$_2$O adsorption, we observed that it adsorbs dissociatively on (001)-(1$\times$1) surface and on ridges of (001)-(1$\times$4) surface, whereas, it is weakly adsorbed without dissociation on (101) and (010) surface. The coadsorption of CO$_2$ and H$_2$O alters the binding energy, and finally lead to formation bicarbonate complex by overcoming energy barriers. The barrier for (001)-(1$\times$1), reconstructed (001)-(1$\times$4), (010) and (101) surfaces are estimated to be 0.44, 0.92, 0.86 and 2.03 eV respectively, which  establishes that (001) surface is most reactive. Even though it reconstructs at high temperature and ultrahigh vacuum, the reactivity towards functional molecules  remains the same, but the number of active sites are reduced by three-fourth. To conclude, the present study explains the site specific reactivity which will be eventually helpful in enhancing the catalytic performance through facet engineering.
\begin{acknowledgement}
This work is supported by Defense Research and Development Organization, India, through Grant No. ERIP/ER/RIC/201701009/M/01. The authors would like to thank HPCE, IIT Madras for providing the computational facility. The authors declare no competing financial interests.
\end{acknowledgement}
\providecommand{\latin}[1]{#1}
\makeatletter
\providecommand{\doi}
  {\begingroup\let\do\@makeother\dospecials
  \catcode`\{=1 \catcode`\}=2 \doi@aux}
\providecommand{\doi@aux}[1]{\endgroup\texttt{#1}}
\makeatother
\providecommand*\mcitethebibliography{\thebibliography}
\csname @ifundefined\endcsname{endmcitethebibliography}
  {\let\endmcitethebibliography\endthebibliography}{}

\end{document}